\documentclass[10pt,conference]{IEEEtran}
\renewcommand{\baselinestretch}{0.980}
\usepackage{multirow}
\usepackage[dvips]{graphicx}
\usepackage{amsmath}
\usepackage{amssymb}
\usepackage{amsxtra}
\usepackage{threeparttable}
\usepackage{cite}
\usepackage{array}
\usepackage{epsfig}
\usepackage{enumerate}
\usepackage{algorithm}
\usepackage{algorithmic}
\usepackage{color}

\begin{document}
%

\title{
A Robust SRAM-PUF Key Generation Scheme Based on Polar Codes 
\vspace{-0.35em}
}

\author{\IEEEauthorblockN{Bin Chen, Tanya Ignatenko, Frans M.J. Willems}
\IEEEauthorblockA{Eindhoven University of Technology\\
		Eindhoven, The Netherlands\\
		Email: \{b.c.chen,  t.ignatenko, f.m.j.willems\}@tue.nl}
\and
\IEEEauthorblockN{Roel Maes, Erik van der Sluis, Georgios Selimis}
\IEEEauthorblockA{Intrinsic-ID,
Eindhoven, The Netherlands\\
Email: \{roel.maes, erik.van.der.sluis,\\ georgios.selimis\}@intrinsic-id.com}
}


%


\maketitle

\begin{abstract}
Physical unclonable functions (PUFs) are relatively new security primitives used for device authentication and device-specific secret key generation. In this paper we focus on SRAM-PUFs. The SRAM-PUFs enjoy uniqueness and randomness properties stemming from the intrinsic randomness of SRAM memory cells, which is a result of manufacturing variations.  This randomness can be translated into the cryptographic keys thus avoiding the need to store and manage the device cryptographic keys. Therefore these properties, combined with the fact that SRAM memory can be often found in today's IoT devices, make SRAM-PUFs a promising candidate for securing and authentication of the resource-constrained IoT devices.
PUF observations are always effected by noise and environmental changes. Therefore secret-generation schemes with helper data are used to guarantee reliable regeneration of the PUF-based  secret keys.  Error correction codes (ECCs) are an essential part of these schemes.
In this work, we propose a practical error correction construction for PUF-based secret generation that are based on polar codes.
The resulting scheme can generate $128$-bit keys using $1024$ SRAM-PUF bits and $896$ helper data bits and achieve a failure probability of $10^{-9}$ or lower for a practical SRAM-PUFs setting with bit error probability of $15\%$.
The method is based on successive cancellation combined with list decoding and hash-based checking that makes use of the hash that is already available at the decoder.
In addition, an adaptive list decoder for polar codes is investigated. This decoder increases the list size only if needed.
\end{abstract}


%
\IEEEpeerreviewmaketitle

 \vspace{-0.2em}
\section{Introduction}
{The Internet of Things (IoT) is a network, in which billions of  devices are connected. While such a network is expected to bring  tremendous economic benefits to industry and society, its use also comes with security problems.
Most of IoT devices operate in resource-constrained and distributed environments.
As a result traditional password-based security and centralized key management systems with costly secure elements cannot be easily deployed in IoT networks. }

{Physical unclonable functions (PUFs)
are  low-cost hardware intrinsic security primitives that possess an intrinsic randomness (unique device `\textit{fingerprint}") due to the inevitable process variations during manufacturing. 
Therefore, PUFs can be used to realize cryptographic applications, such as identification, authentication and cryptographic key  generation \cite{Suh2007,Hussain2016}, that  require random, unique and unpredictable keys.
Since the device-unique randomness can be translated into a cryptographic key, PUFs can act as trust anchors avoiding the need for key storage.}

{There are several types of structures to realize PUFs, such as Flip-Flops PUFs \cite{Maes08intrinsicpufs}, Butterfly PUFs \cite{Kumar08}, Ring Oscillator PUFs \cite{Gassend2002} and static random-access memory (SRAM) PUFs \cite{Guajardo2007}. Among them,  SRAM-PUFs are one of the most popular PUF constructions because they are easy to manufacture and do not require extra investments.}
SRAM-PUFs also enjoy the properties that, while being easily evaluated (after a device power-up),  they are unique, reproducible, physically unclonable and unpredictable \cite{Maes2010}.
However, SRAM-PUFs cannot be straightforwardly used as cryptographic keys, since their observations are not exactly reproducible due to environmental condition changes such as time, temperature, voltage and random noise.
Therefore, error correction techniques are necessary to mitigate these effects and generate reliable keys.

Error correction techniques become essential blocks of secret-generation schemes \cite{Maurer1993, Ahlswede1993,Dodis2004}.   In these schemes two terminals observe measurements of the same PUF. The encoder (first terminal) creates a secret-key and a so-called helper data, based on its PUF observation. This helper data facilitates reconstruction of the secret key from the noisy observation of the PUF at the decoder (second terminal). Since the helper data is communicated from the encoder to the decoder, the secrecy leakage (information that it provides about the secret key) should be negligible.

For practical implementation of key generation schemes on resource-constrained PUF devices, especially for IoT applications, 
it is crucial to construct good error correction codes to maintain a good trade-off between reliability, implementation complexity and secrecy leakage. 
Most of existing works  \cite{Dodis2004,Maes2009,Maes2012,puchinger2015error,Hiller2016} that use simple error correction codes are impractical for real applications, where environmental variations lead to error rates of up to $25\%$ in PUF observations.
These high error rates require (simple) ECCs of low rates. On the other hand, security applications  impose requirements on the minimum (fixed) secret key size that need to be generated from a given finite block-length SRAM cells. As a result, one need to use powerful high-rate ECCs, which typically have high complexity.

Therefore, in this work we propose to use  polar codes that are capacity-achieving and have low encoding and decoding complexity.
Polar codes have been also investigated for the Slepian-Wolf problem \cite{Urbanke2010} and  key generation  \cite{Chou2015}. For finite block-length, it was shown that good performance of polar codes can be achieved by implementing enhanced decoding algorithms based on the classical successive cancellation decoder (SCD) \cite{Arikan2009,Niu2012,Tal2015}.

{Here we propose a new and efficient key generation building block for SRAM-PUFs key generation based on application of polar codes in a syndrome-based secret-generation scheme  \cite{Draper2007}.
To guarantee the performance in terms of reliability and security, and to decrease the required memory size of this scheme, we 
(1) exploit the efficient decoding algorithm based on successive cancellation and list decoding to reliably regenerate the secret,
(2) prove zero-leakage for the proposed scheme,
and (3) use a puncturing scheme to shorten the code length and reduce the complexity.
Our simulation results show that $10^{-9}$ key regeneration failure probability can be achieved with less SRAM-PUF  and helper data bits than before. Using puncturing for polar coding schemes results in flexibility in getting the required code rates, which is crucial since key sizes in practical applications are typically fixed.}

 \vspace{-0.2em}
\section{Secret Generation based on SRAM-PUFs}
\begin{figure}[!t]
  \centering
  \includegraphics[width=3.45in]{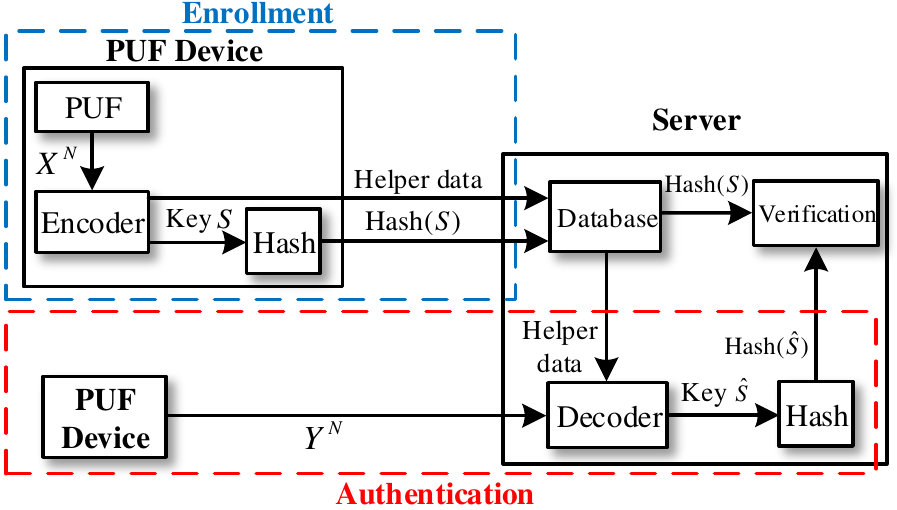}
  \vspace{-2em}
  \caption[]{Secret-generation system for PUFs}
   \vspace{-1.7em}
  \label{fig:model}
\end{figure}

SRAM-PUFs are a result of the read-outs of the power-up state of an SRAM array. The cell values of SRAM array after power up go into one of two states: $0$ or $1$.  It has been experimentally demonstrated \cite{Schrijen2012} that due to the independent random nature of process variations on each SRAM cell, SRAM patterns demonstrates excellent PUF behavior, i.e. empirical probability of number of cells that go in state $1$ is close to $0.5$. Therefore in this paper we assume that  SRAM-PUFs are binary-symmetric, hence for enrollment and authentication PUF pairs $(X^N, Y^N)$ it holds that
\begin{equation}
\Pr\{ (X^N,Y^N) = (x^N,y^N) \} = \prod_{n=1}^N Q(x_n,y_n),
\end{equation}
where $Q(0,1)=Q(1,0)=p/2$ and $Q(0,0)=Q(1,1)=(1-p)/2$ and $0 \leq p \leq 1/2$.

{It is our goal to share a PUF-based secret key $S$ between a PUF-device and a server, see Fig.~\ref{fig:model}.
During the enrollment phase, the encoder observes SRAM-PUF measurement $X^N$ 
and based on it generates a secret key $S$ and helper data $W,$ as $(S,W)=enc(X^N).$}
Here $enc(\cdot)$ is an encoder mapping. Since the key is used for cryptographic purposes, it has to be uniformly distributed.
Moreover, the helper data is assumed to be publicly available, and thus it should leak no information about the key, i.e. $I(S;W)=0$.

Next during the secret regeneration phase, the decoder observes the authentication SRAM-PUF measurement $Y^N$ and the corresponding helper data $W.$ The decoder now forms an estimate of the secret key as $\widehat{S}=dec(Y^N,W),$  with $dec(\cdot)$ being a decoder mapping. To make an authentication decision the server compares the hash of the estimated secret key, $Hash(\widehat{S})$ with $Hash(S).$\footnote{A  one-way cryptographic hash function is used to generate a hash value of the key and verify whether the key is recovered exactly.
The design and security properties of such one-way cryptographic hash functions is beyond the scope of this paper.}
The authentication decision is positive only if  the hashes are the same and thus the secret reconstruction was successful. Hence to ensure the system reliability, the error or failure  probability $\Pr\{\widehat{S} \neq S\}$ should be small.

The secret-generation problem is closely related to the Slepian-Wolf coding problem and is often realized using syndrome construction, where the helper data is the syndrome of the enrollment observation. Due to high error rates in SRAM-PUFs,  $15\%-25\%$, combined with demands of having $\Pr\{\widehat{S} \neq S\}$ of $10^{-9}$ in practical application, powerful codes are required for reliable key generation. 
In this paper we explore the use of polar codes for SRAM-PUF secret generation based on syndrome construction.

 \vspace{-0.5em}
\section{Polar Codes}\label{sec:polar_codes}
As a family of linear block codes,  a binary polar code can be specified by $(N,K,\mathcal{F},u^{\mathcal{F}})$, where $N=2^n$ is the block length,
$K$ is the number of information bits encoded per codeword, $\mathcal{F}$ is a set of indices for the $N-K$ frozen bits positions from $\{1,2,\ldots, N\}$  and $u^{\mathcal{F}}$ is a vector of frozen bits.
The frozen bits are assigned by a fixed binary sequence, which is known to both the encoder and the decoder.

 \vspace{-0.5em}
\subsection{Code Construction of Polar Codes}
Polar codes are channel specific codes, which means that a polar code designed for a particular channel might not have an optimal performance for other channels.
Therefore, calculation of channel reliability and selection of good channels
is a critical step for polar coding, which is often referred to as \emph{polar code construction}. The original construction of polar codes is based on the Bhattacharyya bound approximation \cite{Arikan2009}. Later works  
\cite{Mori2009,Tal2013}
improve on this approximation, however, at the cost of higher complexity.
\vspace{-0.5em}
\subsection{Encoding of Polar Codes}
For an $(N,K,\mathcal{F})$ polar code, the encoding operation for a vector of information bits, $\bf{u}$, is performed using a generator matrix,
\begin{equation}
{\mathbf{G}}_N={\mathbf{G}}_2^{\otimes \log N},
\end{equation}
where ${\mathbf{G}}_2=
\left[
\begin{array}{cc}
   1 & 0 \\
   1 & 1
\end{array}
\right]$
and $\otimes$ denotes the Kronecker product. Given the data sequence ${U}$, the codewords are generated as
\begin{equation}
\begin{aligned}
V=U{\mathbf{G}}_N 
 =U^{\mathcal{F}^c}{(\mathbf{G}_N)}_{\mathcal{F}^c}+U^{\mathcal{F}}{(\mathbf{G}_N)}_{\mathcal{F}},
 \end{aligned}
\end{equation}
where $\mathcal{F}^c \triangleq \{1,2, \ldots,N\}\backslash \mathcal{F}$ corresponds to the non-frozen bits indices. Then $U^{\mathcal{F}^c}$ is the data sequence, and $U^{\mathcal{F}}$ are the frozen bits, which are usually set to zero.

\vspace{-0.5em}
\subsection{Decoding of Polar Codes}\label{sec:decoding}
Polar codes achieve the channel capacity asymptotically in code length, when decoding is done using the successive-cancellation (SC) decoding algorithm, which sequentially estimates the bits $\hat{u}_i$, where $0\leq i \leq N$.

When polar decoder decodes the $i$th bit, $\hat{u}_i$ is estimated based on the channel output $y^N$ and the previous bit decisions $\hat{u}_{1}, \hat{u}_{2},\ldots, \hat{u}_{i-1}$, denoted by $\hat{u}_1^{i-1}$. It uses the following rules:
\begin{equation}\label{eq:PC_SCD}
\hat{u}_i=
\left\{
\begin{aligned}
&u_i, \ \ \ \text{if} \ i\in\mathcal{F}  \\
&0,  \ \ \ \ \text{if} \ i\in\mathcal{F}^c \ \text{and} \ L(y_1^N,\hat{u}_1^{i-1})\geq 1  \\
&1,  \ \ \ \ \text{if} \ i\in\mathcal{F}^c \ \text{and} \ L(y_1^N,\hat{u}_1^{i-1})<1
\end{aligned}
\right.,
\end{equation}
where $L_{N}^{i}(y_1^N,\hat{u}_1^{i-1})= \frac{Pr(0|y_{1}^{N},u_{1}^{i-1})}{Pr(1|y_{1}^{N},u_{1}^{i-1})}$
is the $i$th likelihood ratio (LR) at length $N$,
which determines the probability of a non-frozen bit.
LRs can be computed recursively using two formulas:
\begin{align}\label{eq:LLR_cal}\nonumber
&L_{N}^{2i-1}(y_1^N,\hat{u}_1^{2i-2})\\ 
&=\frac{L_{N/2}^{i}\left(y_{1}^{N/2},\hat{u}^{2i-2}_{o}\oplus \hat{u}^{2i-2}_{e}\right) L_{N/2}^{i}\left(y_{1}^{N/2+1},\hat{u}^{2i-2}_{e}\right) +1}{L_{N/2}^{i}\left(y_{1}^{N/2},\hat{u}^{2i-2}_{o}\oplus \hat{u}^{2i-2}_{e}\right)+L_{N/2}^{i}\left(y_{1}^{N/2+1},\hat{u}^{2i-2}_{e}\right)} 
\end{align}
and
\begin{align}\nonumber
&L_{N}^{2i}(y_1^N,\hat{u}_1^{2i-1})=\left[L_{N/2}^{i} \left(y_{1}^{N/2},\hat{u}^{2i-2}_{o}\oplus \hat{u}^{2i-2}_{e}\right)\right]^{1-2\hat{u}_{2i-1}} \\
 &\ \ \ \ \ \ \ \ \ \ \ \ \ \ \ \ \ \ \ \cdot L_{N/2}^{i}\left(y_{1}^{N/2+1},\hat{u}^{2i-2}_{e}\right),
\end{align}
where $\hat{u}^{2i-2}_{o}$ and $\hat{u}^{2i-2}_{e}$  denote, respectively, the odd and even indices part of $\hat{u}^{2i-2}$.
{Therefore, calculation of LRs at length $N$ can be reduced to calculation of two LRs at length $N/2$, and then recursively broken down to block length 1.
The initial LRs can be directly calculated from the channel observation.}

Since the cost of implementing these multiplications and divisions operations in hardware is very high, they are usually avoided and performed in the logarithm domain using  the following $f$ and $g$ functions:
\begin{eqnarray}\label{eq:f1}
f(L_1,L_2)&= 2\tanh^{-1}\left(\tanh\left(\frac{L_1}{2}\right)\tanh\left(\frac{L_2}{2}\right)\right)\\
&\approx \text{sign}(L_1\cdot L_2)\cdot \min\left(|L_1|,|L_2|\right),\label{eq:f2}
\end{eqnarray}
\begin{equation}\label{eq:g}
g(L_1,L_2)=(-1)^{1-2\hat{u}_{2i-1}}\cdot L_1+L_2,
\end{equation}
where 
$L_1=\log \left( L_{N/2}^{i}\left(y_{1}^{N/2},\hat{u}^{2i-2}_{o}\oplus \hat{u}^{2i-2}_{e}\right)\right)$ and  $L_2=\log \left(L_{N/2}^{i}\left(y_{1}^{N/2+1},\hat{u}^{2i-2}_{e}\right)\right)$ are log-likelihood ratios (LLRs).  
In practical implementations, the \textit{minimum} function can be used to approximate the $f$ function, according to (\ref{eq:f2}).

\vspace{-0.2em}
\section{Secret-Generation Schemes based on Polar Codes}
In this section we show how secrets and helper data can be constructed using a polar code in PUF-based key generation schemes.
{A generic secret-generation system is illustrated in Fig. \ref{fig:model}. There $X^N$ is a PUF measurement during enrollment and  $Y^N$ is a noisy PUF measurement at authentication, which are observed by the encoder and decoder, respectively.
First, in Section \ref{sec:syndrome}, we present the secret-generation system based on syndrome construction using the polar coding. Then, in Section \ref{sec:decopt} we discuss how the decoding for secret generation can be redesigned to optimize the system performance for PUF applications. Finally, Section \ref{sec:security} provides our security analysis for the proposed construction.


\vspace{-0.3em}
\subsection{Polar Codes based Syndrome Construction}\label{sec:syndrome}
Fig. \ref{fig:Polar_codes_syndrome_coding_scheme} illustrates the polar code based syndrome coding scheme that realizes an enrollment phase (encoder) and key regeneration phase (decoder).
\begin{figure}[!t]
  \centering
  \includegraphics[width=3.2in]{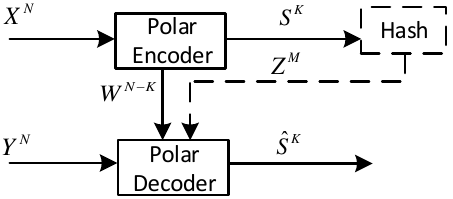}
  \caption{Polar codes based syndrome coding scheme. Dashed line indicate the extra operation for polar code with HA-SCL decoding.}
  \label{fig:Polar_codes_syndrome_coding_scheme}
 \vspace{-1.7em}
\end{figure}
\vspace{-0.1em}
\subsubsection{Enrollment phase}
In the enrollment phase, a codeword $C^N=X^{N}{\mathbf{G}}_{N}^{-1}$ is generated for each PUF observation $X^N$. Then, the syndrome encoder selects the secret key $S^K$ and helper data $W^{N-K}$ based on the constructed codeword. 
{Since $\mathbf{G}_N^{-1}=\mathbf{G}_N$, the helper data and the secret key are generated during a polar encoding procedure by extracting the bits as}\footnote{
The difference compared to conventional polar codes is that the helper data specifies a coset of the linear polar code instead of fixed  all-zeros.}
\begin{equation}\label{eq:generator_help_key}
\begin{aligned}
W^{N-K} &\triangleq \left(X^{N}{\mathbf{G}}^{-1}_{N}\right)_{\mathcal{F}}=\left(X^{N}{\mathbf{G}}_{N}\right)_{\mathcal{F}}=C^{N}[\mathcal{F}],\\
S^K &\triangleq \left(X^{N}{\mathbf{G}}^{-1}_{N}\right)_{\mathcal{F}^c}=\left(X^{N}{\mathbf{G}}_{N}\right)_{\mathcal{F}^c}=C^{N}[\mathcal{F}^c],
\end{aligned}
\end{equation}
where $\mathcal{F}$ and $\mathcal{F}^c$ are the index sets for the syndrome and the secret key. These sets are defined as
\begin{equation}
\begin{aligned}
\mathcal{F}&\triangleq  \left\{i\in \{1,2,\ldots, N\}: H(C_i|Y^{N},C^{i-1}_1)\geq \delta\right\}, \\
\mathcal{F}^{c}&\triangleq \left\{1,2,\ldots, N\right\} \backslash \mathcal{F},
\end{aligned}
\end{equation}
where $\delta\triangleq 2^{-N\beta}$ and $\beta \in [0,1/2]$.

An example of the syndrome encoding procedure for a (8,3,\{1,2,3,4,6\}) polar codes is shown in Fig. \ref{fig:Polar_codes_encoding}, where the data flows from right to left.
Due to flexibility of the polar code construction, an arbitrary code rate $R=K/N$ can be selected without re-constructing the code.
\begin{figure}[!t]
\centering
  \includegraphics[width=3.3in]{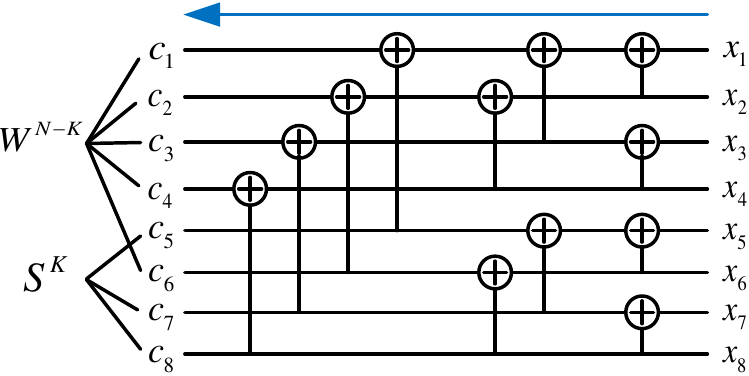} 
  \caption{Encoding graph of $\left(8,3,\{1,2,3,4,6\}\right)$ polar codes}
  \label{fig:Polar_codes_encoding}
  \vspace{-0.0em}
\end{figure}
  \vspace{-0.1em}
\subsubsection{Key regeneration phase}
In the key regeneration phase, the syndrome decoder observes the authentication sequence $Y^N$  and also receives the public helper data $W^{N-K}$. The decoder can compute $\hat{S}^K$ using a modified version of the SC decoding algorithm, given in Section \ref{sec:decoding}, i.e.,  as
\begin{equation}\label{eq:SC}
\hat{S}^{K}=\text{SCD} (Y^{N},W^{N-K}),
\end{equation}
where the polar decoder $\text{SCD}(\cdot)$ is given by Algorithm \ref{alg:Syndrome_construction}.

\begin{algorithm}
 \caption{Decoding Algorithm for Syndrome construction}
 \begin{algorithmic}[1]\label{alg:Syndrome_construction}
 \renewcommand{\algorithmicrequire}{\textbf{Input:}}
 \renewcommand{\algorithmicensure}{\textbf{Output:}}
 \REQUIRE The observations $Y^{N}$ from PUFs, the public helper data $W^{N-K}$.
 \ENSURE  The estimated secret $\hat{S}^K$
  \FOR {$i = 1$ to $N$}
  \STATE Compute $LLR_i$ with the observed $Y^{N}$ from Eq. (\ref{eq:f1}-\ref{eq:g})
  \IF {$i \in \mathcal{F}$}
  \STATE $\hat{C}_i=W_j$
  \ELSIF {$i\in \mathcal{F}^c$ and $LLR_i>0$}
  \STATE $\hat{C}_i=0$
  \ELSE
  \STATE $\hat{C}_i=1$
  \ENDIF
  \STATE $\hat{S}^K \leftarrow \hat{C}^N[\mathcal{F}^c]$
  \ENDFOR
 \RETURN $\hat{S}^K$ 
 \end{algorithmic}
 \end{algorithm}

\subsection{Decoder Optimization for PUF-based Secret Generation}\label{sec:decopt} 
Note that although the SC decoder could asymptotically achieve channel capacity as $N$ increases, the performance of the SC decoder is still not good enough at short and moderate block length size for error correction in PUFs due to the poor polarization.
Therefore, next we present hash-aided SC list (HA-SCL) decoding that allows us to achieve good trade-off between error-correction performance and complexity.

In order to  optimize the error-correction performance, we would like to track multiple possible decision paths instead of only one as the SC decoder does. However, considering the all $2^{K}$ possible paths is impractical and too complex. 
The SCL decoding algorithm \cite{Tal2015} uses a breadth search method to explore the possible decoding paths efficiently while saving  $L$ most reliable paths as candidates at each level. Thus this technique also allows us to restrict the decoding complexity.

Next note that in the SCL decoding process, the correct codewords are on the decoding list but they are not always the most likely ones, which leads to decoding errors. This issue can be solved by combining the SCL algorithm with a cyclic redundancy check (CRC) code, which could further improve the error correction performance \cite{Niu2012}.
{For security purposes,  we replace the CRC function by a  more secure hash function, which is already part of the authentication system. This hash is used to detect and select the valid path from the output of list decoder.}
In this way, our HA-SCL decoder outputs $L$ candidate sequences and selects the hash-valid sequence.

{By using the HA-SCL decoder, $M$ bits hash value $Z^M=H(S^K)$ is produced by the hash function at the encoder and is used at the decoder.}
Since the decoder knows $W^{N-K}$ and $Z^{M}$ in advance, it could recover the secret key by performing the polar decoding, as shown in Fig. \ref{fig:Polar_codes_syndrome_coding_scheme}, using
\begin{equation}\label{eq:SCLD}
\hat{S}^{K}=\text{SCLD} (Y^{N},W^{N-K},Z^{M}),
\end{equation}
where $\text{SCLD}(\cdot)$ is the polar decoder with the SCL decoding algorithm of \cite{Tal2015}.

\vspace{-0.3em}
\subsection{Security Analysis}\label{sec:security}
In this section, we analyze the secrecy for the proposed syndrome based polar coding scheme.
Note that security of our construction is characterized by the information that the helper data leaks about the generated secret key. Therefore we must show that $I(S^K;W^{N-K})=0$. 
We re-write (\ref{eq:generator_help_key}) as
\begin{equation}
\begin{aligned}
W^{N-K}&= (X^N \mathbf{G})_{\mathcal{F}}=X^N \mathbf{G}_{\mathcal{F}},\\  
S^K&= (X^N \mathbf{G})_{\mathcal{F}^c}=X^N \mathbf{G}_{\mathcal{F}^c},
\end{aligned}
\end{equation}
where  generator matrix ${\mathbf{G}}_\mathcal{F}$  for frozen bits (helper data) and generator matrix ${\mathbf{G}}_{\mathcal{F}^c}$  for information bits (key) with  dimensions $N\times (N-K)$ and $N\times K$ are obtained by selecting the corresponding columns of ${\mathbf{G}}_N$.
Then, we obtain
\begin{equation}
\begin{aligned}\nonumber
I(S^K;W^{N-K})&=H(S^K)+H(W^{N-K})-H(S^K,W^{N-K})\\
      &=H(X^N{\mathbf{G}}_{\mathcal{F}^c})+H(X^N{\mathbf{G}}_\mathcal{F})-H(X^N\mathbf{G})\\
      &\overset{a}{=}rank({\mathbf{G}}_{\mathcal{F}^c})+rank({\mathbf{G}}_\mathcal{F})-rank({\mathbf{G}})\\
      &\overset{b}{=}K+(N-K)-N\overset{}{=}0,
\end{aligned}
\end{equation}
where in (a) we use the  uniformity of the SRAM-PUF observations; in (b) the fact that the generator matrices are linearly independent, as ${\mathbf{G}}_\mathcal{F}$ and ${\mathbf{G}}_{\mathcal{F}^c}$ are non-overlapping, since $\mathcal{F}^c \cap \mathcal{F}=\phi$. Thus we prove that the proposed polar syndrome coding scheme has zero-leakage.
\vspace{-0.2em}
\section{Performance and Complexity Comparisons}
In this section, we present the performance results of the polar code based error correction schemes for SRAM-PUFs with average bit error probability between $15\%$ and $30\%$.
Inputs to the polar decoder, including the information set, frozen bit vector and channel output vector, determine the error correction ability and computational complexity.
In order to create reliable PUF-based secret generation systems, we focus on the scheme with 128-bit keys and failure probabilities  in the range of $10^{-6}$ to $10^{-9}$.

We construct polar codes with block length $N=1024$.
In order to provide a flexible code rate and use less SRAM-PUFs bits in PUF-based secret generation with fixed size key, arbitrary block-length polar codes can be obtained by puncturing.
For any puncturing pattern, $N'=N - m$ PUFs bits and $m$ random bits used as punctured bits are the input to the polar encoder.
At the decoder, $m$ zero-valued LLRs for decoding are assigned to the corresponding punctured bits.
In the following sections, both SC and HA-SCL decoding algorithms for polar codes and punctured polar codes are simulated to compare the resulting error correction performance and complexity.
\begin{figure}[!t]
  \centering
  \includegraphics[width=3.in]{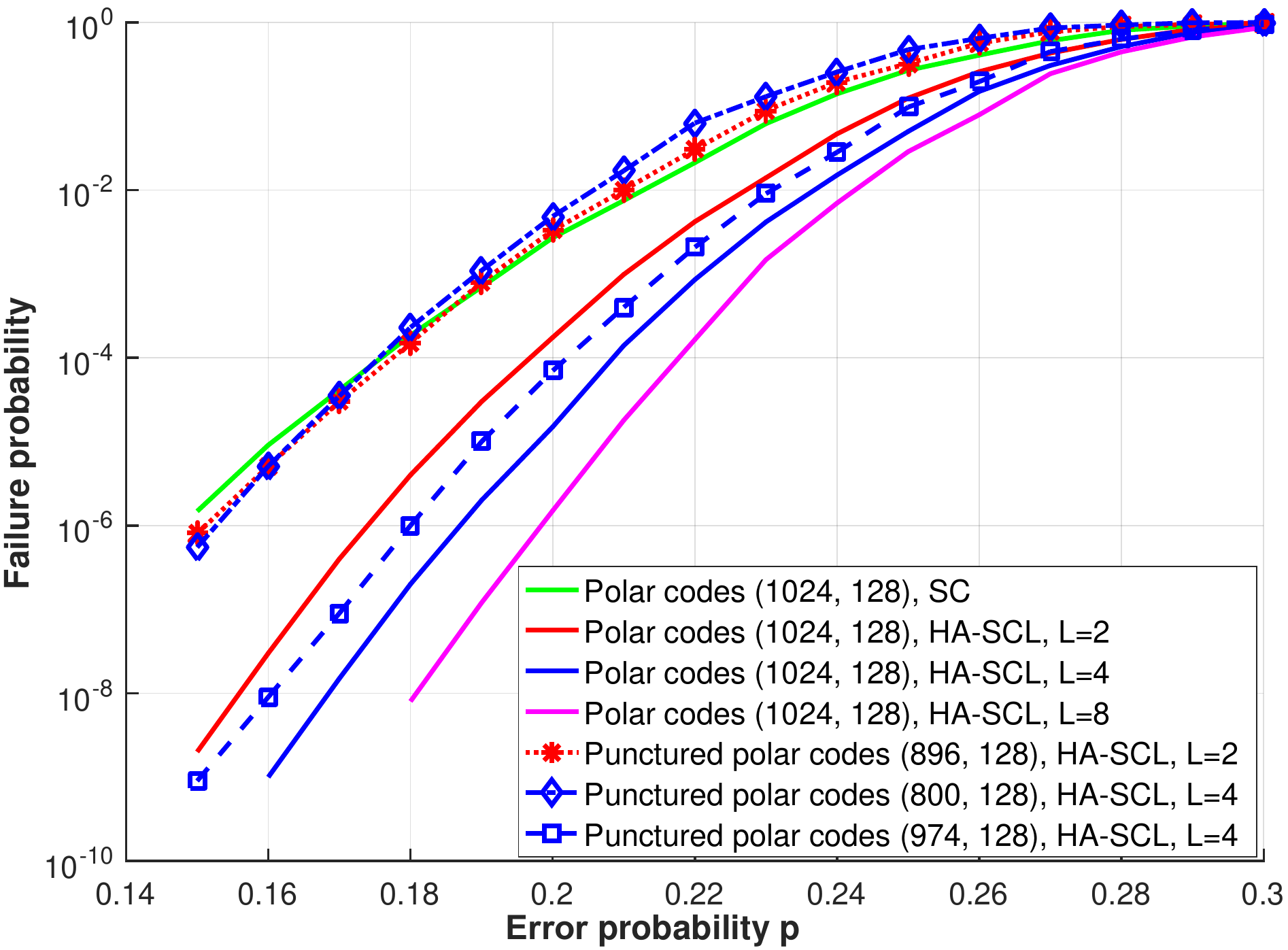}
  \caption{Failure rate performance comparison of the different decoding schemes.}
  \label{fig:simulation_polar}
   \vspace{-1.5em}
\end{figure}

 \vspace{-0.3em}
\subsection{Failure Probability}
The most important performance criterion for PUF-based secret generation is the error or failure probability of the key regeneration.
Fig. \ref{fig:simulation_polar} shows the performance of polar code based syndrome coding schemes with the SC and HA-SCL decoding algorithms. 

We can see that the failure rate for polar codes with SC decoding is close to $10^{-6}$ at $15\%$ and  HA-SCL decoding can further reduce the failure rate to less than $10^{-9}$ at $15\%$ as list size $L$ increases. However, the latter comes at the cost of extra computational complexity and memory.
We can also observe that punctured polar codes with block lengths not being a power of two achieve similar failure rates as conventional polar codes by using larger $L$ for performance compensation.
For strong reliability applications, the proposed polar code $(1024, 128)$ with $L=2$ and the punctured polar code $(974, 128)$ with $L=4$ can be used to achieve an error rate of $10^{-9}$ at error probability of $15\%$.
\begin{table*}[!t]
\centering
\caption{Comparison of failure rate, PUF and helper data size for polar codes and reference designs. }
\vspace{-0.8em}
\label{tab:comparison}
\begin{tabular}{l|l|c|c|c}
Code construction        & Failure probability                          & PUF (bit)  & Helper Data (bit)& $p$  \\ \hline  \hline
Code-Offset RM-GMC\cite{Maes2009} & $10^{-6}$           & 1536              & 13952       & 15\% \\ \hline
Compressed DSC \cite{Hiller2016}    & $10^{-6}$           & 974               & 1108        & 15\% \\ \hline
Polar SC                 & $10^{-6}$         & 1024              & 896         & 15\% \\ \hline
Punctured polar HA-SCL, L=2                 & $10^{-6}$         & 896              & 896         & 15\% \\ \hline
Punctured polar HA-SCL, L=4                 & $10^{-6}$         & 800              & 896         & 15\% \\ \hline  \hline
BCH Rep. \cite{Maes2012}          & $10^{-9}$           & 2226             & 2052        & 13\% \\ \hline
GC RM \cite{puchinger2015error}             & $5.37 \cdot 10^{-10}$  & 2048              & 2048        & 14\% \\ \hline
GC RS \cite{puchinger2015error}        & $3.47 \cdot 10^{-10}$ & 1024              & 1024        & 14\% \\ \hline
Polar  HA-SCL, L=2           & $10^{-9}$           & 1024              & 896         & 15\% \\ \hline
Punctured polar HA-SCL, L=4                 & $10^{-9}$         & 974              & 896         & 15\% \\ \hline
\end{tabular}
\vspace{-1.9em}
\end{table*}

\vspace{-0.3em}
\subsection{Complexity and Memory Requirements}
The number of required SRAM-PUF bits and helper data size is another important performance criterion closely related to implementation complexity.
The proposed polar code based syndrome coding scheme requires $1024$ SRAM-PUF bits and $896$ helper data bits, if we  implement the SC decoding algorithm. The corresponding decoding computational complexity is given by $\mathcal{O}(N\log N)$ for this case.
Since a list decoder outputs a group of $L$ reliable candidates,
the decoding computational complexity of the HA-SCL algorithm increases to $\mathcal{O}(LN\log N)$.

Table \ref{tab:comparison} summarizes performance properties for the proposed polar codes and reference designs, including the achievable key regeneration failure rate, the required SRAM-PUF size, helper data size and error probability $p$ of the SRAM-PUFs.
An SRAM-PUF with an error probability of  $15\%$ or lower and failure rates $10^{-6}$ and $10^{-9}$ are targeted for different use cases.
From Table \ref{tab:comparison}, we can clearly see that our polar code based schemes outperform the  previous designs in terms of the error correction performance, SRAM-PUF bits and helper data bits requirements.
{Note that the required SRAM-PUF size could be further reduced by using punctured polar codes and  increasing the list size $L$, but at the cost of computational and memory complexity for decoder.}

\vspace{-0.4em}
\subsection{Adaptive Decoder}
The HA-SCL decoding algorithm achieves a good performance but has higher complexity $\mathcal{O}(LN\log N)$ than SC decoding, as $L$ and $N$ increase, and as a consequence relatively high latency.
The complexity issue of the SCL can be improved by using an adaptive decoder, which consists of two components, SC and SCL decoders.
This adaptive decoder, only implements the SCL decoder and increases the value of $L$, when the SC decoder output has an invalid hash vector.

\begin{figure}[!t]
  \centering
  \includegraphics[width=3.2in]{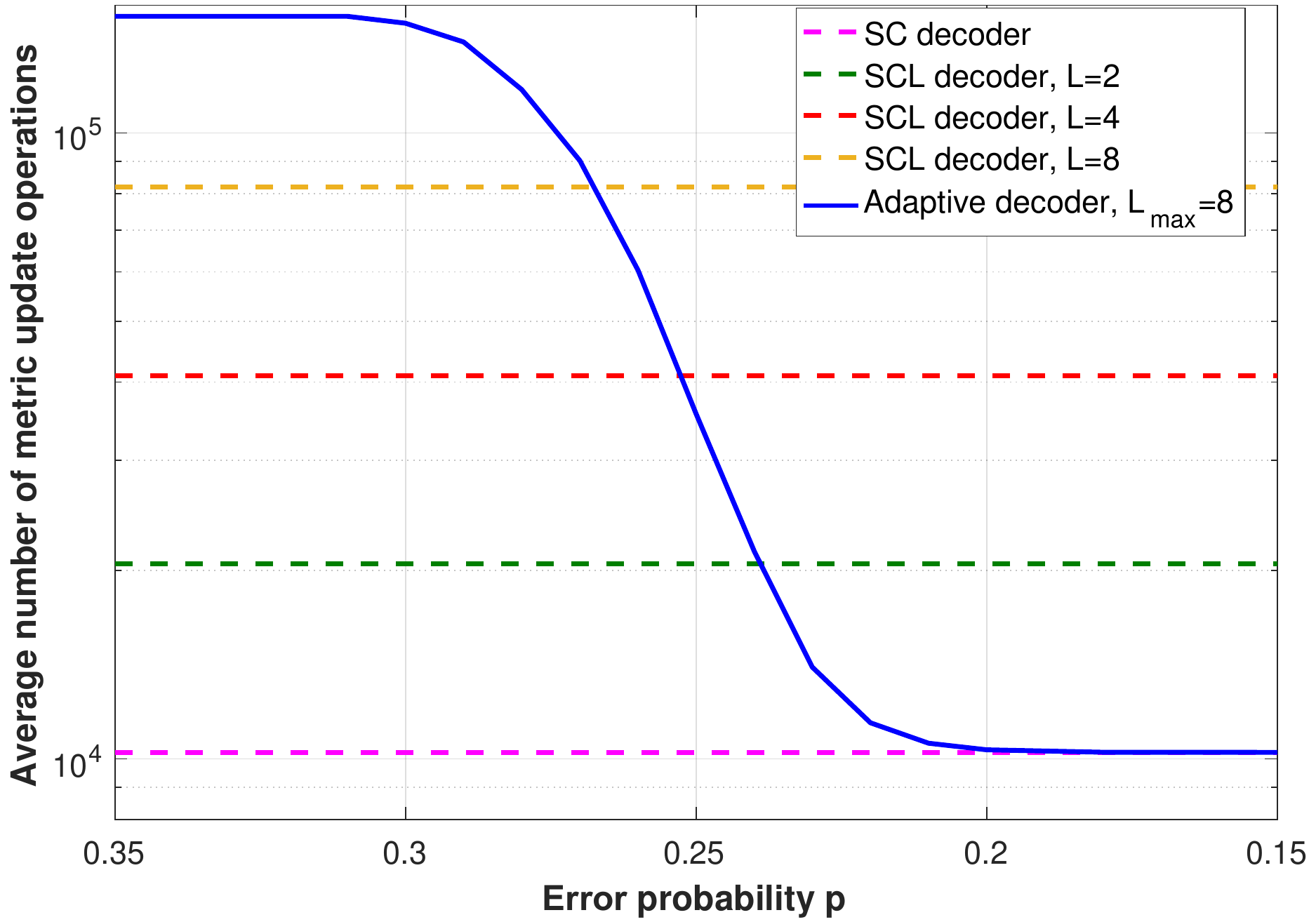}
  \caption{Complexity comparison of the different polar decoding schemes.}
  \label{fig:simulation_polar_comlexity}
  \vspace{-1.5em}
\end{figure}
Fig. \ref{fig:simulation_polar_comlexity} shows the comparison of complexity between the adaptive decoder, single SC decoder and single HA-SCL decoder with different $L$.
Computational complexity is defined in terms of  the average number of metric update operations $f(\cdot)$ and $g(\cdot)$ in (\ref{eq:f1}) and (\ref{eq:g}). The maximum $L_{max}$ is set to $8$ to  ensure the reliability and security.
As expected,  the SC decoder  has the lowest complexity with poor performance with respect to the failure rate; and the SCL decoder has higher complexity in terms of $L$.
Furthermore, we see that the average number of computations of the adaptive decoder is drastically reduced as the error probability $p$ decreases. 

{The adaptive decoder also reduce the effect of decoding latency, since there is very little chance to use the complex SCL decoder.
Therefore, the adaptive decoder could achieve the same reliability with a single SCL decoder and  provide similar computational complexity and decoding latency with a single SC decoder when $p$ is small.}
\vspace{-0.3em}
\section{Discussion}
Note that another way to realize PUF-based authentication is using the \textit{code-offset construction}. In this construction, a selected error correction codes is used to encode a key chosen during the enrollment phase  into codeword $C^N=S\mathbf{G}$. The helper data $W^N$ is defined as the offset $W^N=C^N\oplus X^N$. 
During the key regeneration phase, the helper data $W^N$ is added to a PUF authentication sequence $Y^N$. The decoder observe a codeword corrupted with the measurement noise $e^N$, i.e., $\widetilde{C}^N=W^N\oplus Y^N =C^N \oplus e^N$.
Therefore, polar codes based code-offset construction can also be directly applied to realize a secret-sharing system with chosen secret keys.
 
By designing the same polar code construction, the two secret key generation schemes with the syndrome construction and code-offset construction are equivalent in terms of error correction performance but they differ in their helper data storage requirements. In particular,  the syndrome construction requires less storage for the helper data.
Moreover, the proposed polar codes based  syndrome coding construction potentially has more applications, since the secret keys need not be known to the manufacturer, while the code-offset construction requires the key to be assigned to the PUF devices during the manufacturing process.

\vspace{-0.3em}
\section{Conclusion}
\vspace{-0.2em}
In this paper, we investigated practical secret-generation schemes based on polar code with syndrome construction that treat the SRAM-PUF observations as a codeword of a polar code and generate helper data as a syndrome of SRAM-PUFs using frozen bits of the polar code. 
Our simulation results show that with this approach high secret generation reliability can be achieved together with high security.
Furthermore, the proposed scheme requires less SRAM-PUF bits and helper data bits compared to existing schemes, which leads to the reduction in memory requirements.

The proposed scheme has higher complexity requirements on hardware than simple algebraic codes used in the previous schemes. 
Therefore it can be used in the scenarios for secret key generation between small IoT devices and servers, which have sufficient resources for decoding.
For future work, we intend to investigate the techniques for encoder and decoder optimization to  further reduces the complexity of decoding thus making it also suitable for small IoT devices.

\vspace{-0.3em}
\section*{Acknowledgment}
\vspace{-0.3em}
This work was funded by  Eurostars-2 joint programme with co-funding from the EU Horizon 2020 programme  under the E! 9629 PATRIOT project.
\vspace{-0.3em}
\bibliographystyle{IEEEtran}
\bibliography{IEEEabrv,bibfile}
\end{document}